# Origin of Operating Voltage Increase in InGaN-based Light-emitting Diodes under High Injection: Phase Space Filling Effect on Forward Voltage Characteristics

Dong-Pyo Han[1, 2, a)], Jong-In Shim[2], Dong-Soo Shin[3]

[1]*Faculty of Science and Technology, Meijo University, 1-501 Shiogamaguchi, Tempaku-ku, Nagoya 468-8502, Japan*

[2]*Dept. of Electronics and Communication Engineering, Hanyang University, ERICA Campus, Ansan, Gyeonggi-do 426-791, Korea*

[3]*Dept. of Applied Physics and Dept. of Bionanotechnology, Hanyang University, ERICA Campus, Ansan, Gyeonggi-do 426-791, Korea*

**Abstract**

As an attempt to further elucidate the operating voltage increase in InGaN-based light-emitting diodes (LEDs), the radiative and nonradiative current components are separately analyzed in combination with the Shockley diode equation. Through the analyses, we have shown that the increase in operating voltage is caused by phase space filling effect in high injection. We have also shown that the classical Shockley diode equation is insufficient to comprehensively explain the *I-V* curve of the LED devices since the transport and recombination characteristics of respective current components are basically different. Hence, we have proposed a modified Shockley equation suitable for modern LED devices. Our analysis gives a new insight on the cause of the wall-plug-efficiency drop influenced by such factors as the efficiency droop and the high operating voltage in InGaN LEDs.

[a)] E-mail: han@meijo-u.ac.jp

Recently, the InGaN-based light-emitting diodes (LEDs) covering the visible and ultraviolet spectral regions have been hailed as one of the great inventions of the twentieth century with the Nobel prize awarded to the three principle innovators, I. Akasaki, H. Amano, and S. Nakamura.[1] Owing to the progress, general lighting based on LEDs has become a reality with the advantages of brightness, energy efficiency, and environment friendliness.[2]

Although the performances of InGaN-based LEDs have significantly improved so far, higher light output power and lower electrical input power, namely higher wall-plug efficiency (WPE) are still required for further replacement of conventional lighting source. However, at high driving currents required for general lighting, the WPE of InGaN-based LEDs is limited by well-known degradation mechanisms such as the efficiency droop and high operating voltage.[3-4] Many physical mechanisms have been proposed to explain the origin of efficiency droop including the hot electrons to the vacuum level by Auger recombination,[5] carrier leakage caused by asymmetry of carrier concentration and mobility,[6] carrier overflow assisted by the piezoelectric field,[7] and phase space filling effect triggering the carrier spill-over.[8] These researches have pointed out that the carrier overflow or leakage from the active region is responsible for the efficiency droop. On the other hand, the origin of high operating voltage of InGaN-based LEDs is typically explained by high series resistance caused by the high activation energy of acceptors in GaN,[9] current crowding,[10] and the contact between metal and semiconductor of the LED device.[11] Still, the physical origin of the degradation mechanisms of WPE is not clarified yet.

It is generally known that the total driving current vs. applied voltage (*I-V*) curve of the *pn*-junction semiconductor device is the most important electrical characteristics and is well explained by the Shockley diode equation.[12-13] The parameters in the Shockley equation such as the ideality factor, the reverse saturation current, and the series resistance represent the physical mechanisms including the carrier transport and recombination in the LED device. In other words, we can infer the problem and remedy it from the information of extracted parameters in the Shockley equation fitted with experimental results. The theory of the Shockley equation was first developed for homojunction *pn* devices based on Si and Ge. In particular, it is deeply related to the diffusion current component in the *pn* junction and is therefore usually referred to as the "diffusion current theory".[13] However, unlike in the homojunction devices, various current transport and recombination mechanisms, not just the diffusion, play important roles in heterojunction-based modern LED devices.[14-16] Hence, the Shockley diode equation is often insufficient to comprehensively explain the *I-V* curves of modern LED devices employing multiple quantum wells (MQWs) as active layers.

In the light-generating LED devices, unlike other semiconductor devices such as photodiodes and solar cells, the injected current can be separated into two components, the radiative current, $I_R$, and the nonradiative current, $I_{NR}$, while the Shockley diode equation only explains the diffusion current. While there have been frequent investigations and analyses of total driving current *I* vs. applied voltage *V* by the Shockley equation,[16-18] the data of $I_R$ and $I_{NR}$ vs. *V* still requires a deeper understanding for the LED device.

In this letter, we aim to understand the origin of WPE degradation mechanisms in InGaN-based LED, the characteristics of $I_R$ and $I_{NR}$ as a function of *V* is precisely investigated. To separate the current components, we utilize the information of the *I-V* curve combined with the IQE characteristics of the LED device. We then carefully analyze each current component as a function of *V* with the Shockley equation. Based on these analyses, we give an explanation on the characteristics of $I_R$ and $I_{NR}$ as a function of *V* including physical mechanisms of carrier transport and recombination. We also show the close interrelationship between efficiency droop and high operating voltage in InGaN-based LED device.

For experiments, a commercial blue LED sample with InGaN MQWs was utilized. The peak wavelength and the chip size of the sample were 450 nm and 600×1250 μm$^2$, respectively. The sample had been fabricated into a chip with lateral electrodes and mounted on a surface-mounted-device (SMD) package without epoxy dome. The *I-V* characteristics were measured by using a Keithley semiconductor parameter analyzer under the pulsed-voltage condition. The light output power was collected by a Si photodiode under the pulsed-current driving condition (pulse period: 100 μs, duty cycle: 1%) to avoid the self-heating effect. The IQE and light-extraction efficiency (LEE) data were obtained by the conventional temperature-dependent electroluminescence (TDEL) measurement, which is widely used in various research works.[19-21]

Figures 1 (a) and (b) depict the IQE values of the sample under investigation as a function of driving current on linear and semi-log scales at cryogenic and room temperature, respectively. The peak value of IQE is continuously increased as decrease in operating temperature from 300K to 25K. Thus, we can think that the measured IQE value at room temperature is reliable. The IQE of the sample reaches a maximum value of ~84% at a low current density of ~3 A/cm$^2$ and decreases. The LEE is estimated as ~64% by using the experimental data of the IQE and the external quantum efficiency (EQE). These are considered as typical IQE, LEE, and EQE characteristics obtained from commercial InGaN-based blue LEDs.[20-22]

Now, from the definition of the IQE, $I_R$ and $I_{NR}$ can be separated from the total current $I$ as expressed in eqs. (1) and (2), respectively:[23]

$$I_R = \eta_{IQE} \cdot I \tag{1}$$

$$I_{NR} = (1 - \eta_{IQE}) \cdot I \tag{2}$$

where $\eta_{IQE}$ is the IQE.

Shown in Fig. 2 is the ideality factors $n_R$ and $n_{NR}$ corresponding to $I_R$ and $I_{NR}$ as a function of $V$, obtained by differentiating $I_R$ and $I_{NR}$ as expressed in eqs. (3) and (4), respectively:[24]

$$n_R = \frac{q}{k_B T} \left( \frac{\partial \ln I_R}{\partial V} \right)^{-1} \tag{3}$$

$$n_{NR} = \frac{q}{k_B T} \left( \frac{\partial \ln I_{NR}}{\partial V} \right)^{-1} \tag{4}$$

where $q$ is the elementary charge, $k_B$ is the Boltzmann constant, and $T$ is the absolute temperature. It is seen that $n_R$ remains at 1 in the low voltage < ~2.45 V and then increases with increasing driving current. On the other hand, $n_{NR}$ remains at 2 in the similar low voltage region and then increases with increasing driving current. The increase in ideality factors at high voltages is caused by the additional potential drop unaccounted for in eqs. (3) and (4). The ideality factor of $I_R$ has been shown to be 1 for the direct band-to-band recombination in MQWs of diffusion carrier.[23] On the other hand, the ideality factor of 2 is predicted to be resulted from two mechanisms, namely, the Sah-Noyce-Shockley (SNS) current and the current under high-level injection condition.[25] In our case, $n_{NR}$ of 2 is a result from the SNS current since it is observed in the low-bias region. Thus, $I_{NR}$ is the result of the nonradiative recombination via defects. The value of $n_{NR}$ greater than 2 in the lower voltage region (< ~2.25 V) is considered to be caused by the tunneling mechanism resulting from the threading dislocations introduced during the epitaxial growth, which seem to saturate in the early stage of driving current.[26]

The additional potential drop can be obtained by comparing the experimental results with the ideal diode behavior.[27] To find the potential drop for each current component, we show $I_R$ and $I_{NR}$ vs. $V$ plotted on semi-log scales in Fig. 3 (a) and (b), respectively. The ideal diode behaviors for $I_R$ and $I_{NR}$, which are plotted as solid lines in Fig. 3 (a) and (b), are expressed in eqs. (5) and (6):

$$I_{R,ideal} = I_{S,R} \exp\left(\frac{qV}{k_B T}\right) \tag{5}$$

$$I_{NR,ideal} = I_{S,NR} \exp\left(\frac{qV}{2k_B T}\right) \tag{6}$$

where $I_{S,R}$ ($I_{S,NR}$) is the reverse saturation current corresponding to $I_R$ ($I_{NR}$). Note the ideality factors of 1 for $I_R$ and 2 for $I_{NR}$ are from the result in Fig.2. As mentioned before, the discrepancy between the ideal curve and experimental data of $I_{NR}$ in the low-bias region (< 2.25 V) is thought to be caused by the tunneling current.[26] The additional potential drop ($\Delta V_R$, $\Delta V_{NR}$) between the ideal curves and the experimental results of $I_R$ and $I_{NR}$ can indicate the different current conduction mechanisms for respective current components.

In aspect of $I_R$, from the Shockley theory,[13] $I_{R,ideal}$ vs. $V$ characteristic in eq. (5) is the current flowing under carrier recombination rate is greater than carrier diffusion (injection) rate, namely, $I_{R,ideal}$ is limited by carrier diffusion rate. The carrier diffusion rate exponentially increase as a function of $V$, while radiative recombination rate is saturated due to phase space filling effect under high injection.[28,29] In this case, carrier injection rate exceed the carrier recombination rate and therefore, $I_R$ is limited by radiative recombination rate.

In rate equation model, the relation between $I_R$ and carrier density can be written as,[28]

$$I_R = qV_{eff}\left[Bnp/(1+\alpha n)\right] \tag{7}$$

where $V_{eff}$, $B$, $n$, and $p$ are effective active volume, bimolecular radiative recombination coefficient, electron and hole density in active region, respectively. $B$ is divided by factor, $(1+\alpha n)$ to include phase space filling effect in high current density.[28,29] Under high-level injection condition, $n$ and $p$ has similar quantity, i.e., $n \approx p$. In low injection, i.e., $\alpha n \ll 1$, eq. (7) can be rewritten since $n$ and $p$ depend on quasi-Fermi level as following:[23]

$$I_R = qV_{eff}Bn_i^2\left[\exp\left(\frac{qV}{k_BT}\right)\right] \quad (8)$$

where $n_i$ is the intrinsic carrier density. $I_R$ in eq. (8) seems to have identical voltage dependence with eq. (5). Thus, $I_R$ can be considered as diffusion limited current under low current injection. However, under high injection, the $I_R$ can be rewritten as following:

$$I_R = qV_{eff}Bn_i^2\left[\exp\left(\frac{qV}{k_BT}\right)\right]\frac{1}{1+\alpha n} \quad (9)$$

Eq. (9) implies that the current flowing is decreased due to phase space filling effect, namely additional voltage drop caused by phase space filling effect ($\Delta V_{PSF}$) should be considered to account for $I_R$ vs. $V$ curve. Since $n$ is proportional to $I_R$[13] and $\Delta V_R$ is composed of $\Delta V_{PSF}$ and conventional ohmic potential drop, i.e., $\Delta V_R = \Delta V_{PSF} + IR_S$, the relation between $\Delta V_{PSF}$ and $I_R$ can be expressed as following:

$$\exp^{-1}(\Delta V_{PSF}) = (1+\alpha' I_R)^{-1} \quad (10)$$

where $\alpha'$ is the parameter corresponding to $\alpha$. Thus, it seems that $\Delta V_R$ is significantly influenced by the degree of $\alpha$, namely higher value of $\alpha$ lead to higher operating voltage. It is note that the $\alpha$ is the parameter representing reduced recombination probability of e-h pair in $k$-space due to difference in effective mass between hole and electron as well as carrier accumulation in reduced effective volume caused by quantum-confined stark effect (QCSE), asymmetry carrier distribution, and local potential fluctuation.[28,30,31] In Fig. 4, $\exp(\Delta V_{PSF})$ vs. $I_R$ curve is plotted and data is fitted with eq. (10). The extraction method of value of $R_S$ is discussed in next section in detail. It is seems that data is fitted very well with theoretical expectation, implying that the additional potential drop $\Delta V_R$ is described by $\Delta V_{PSF} + IR_S$. Hence, we can conclude that phase space filling effect lead to additional potential drop, $\Delta V_{PSF}$, which is comparable to potential drop caused by constant series resistance under high injection.

In aspect of $I_{NR}$, the non-radiative recombination rate in active region is saturated due to such factors as the saturation of the carrier capture rate in defect[32] and phase-space filling effect on Auger recombination rate, i.e., $Cn^3/(1+\alpha n)$ under high injection currents.[28, 29] Consequently, it is induced the condition of carrier injection rate exceeding the recombination rate in the active region. Under this condition, carriers begin to accumulate and transport to retain charge neutrality as a form of drift current. The relation between current and voltage in such drift current has been predicted as the Lampert-Rose law under the name of so-called "double-injection current" in semiconductors:[33,34]

$$J = \frac{9}{8} q (n_0 - p_0) \mu_n \mu_p \tau \left( \frac{V^2}{L^3} \right), \qquad (11)$$

where $n_0$ ($p_0$) is the equilibrium electron (hole) concentration, $\tau$ is the average lifetime of injected carriers, $\mu_n$ ($\mu_p$) is the electrons (holes) mobility, and $L$ is the anode-cathode spacing. In particular, double-injection current has quadratic dependence on the applied voltage. Thus, we can think that $\Delta V_{NR}$ is composed of conventional ohmic potential drop and potential drop by double injection current, i.e., $IR_S + \sqrt{I_{NR}} D_{DI}$. In Fig. 5, the experimental data of $\Delta V_{NR}$ is plotted with $I_{NR}$ and data is fitted with $IR_S + \sqrt{I_{NR}} D_{DI}$. It is seems that data is fitted very well with theoretical expectation, implying that the additional potential drop $\Delta V_{NR}$ can be described by $\Delta V_{NR} = \Delta V_{DI} + IR_S$. In other words, the $I_{NR}$ has a quadratic dependence on the $\Delta V_{NR}$ and $I_{NR}$ is limited by the double injection current, which appears as an additional potential drop in the LED device. In the case of InGaN-based LEDs, it is frequently pointed out that carrier spill-over and overflow from the active region initiated by such factor as the saturation of the recombination rate caused by phase-space filling in the small effective active volume as a form of drift current is as origin of efficiency droop.[6,16,35] We believe that this double-injection current behavior can be considered as another indication that the carrier overflow and spill-over from the active region does occur under high injection. Therefore, we can conclude that the low degree of carrier recombination rate due to phase space filling compared to injection rate can lead to both the efficiency droop and the high operating voltage in InGaN-based LED devices.

Now, we can represent $I_R$ and $I_{NR}$ by using the above information in the Shockley equations as expressed below:

$$I_R = I_{S,R} \exp \left[ \frac{q(V - IR_S - \ln(1 + \alpha' I_R))}{k_B T} \right], \qquad (12)$$

$$I_{NR} = I_{S,NR} \exp \left[ \frac{q(V - IR_S - \sqrt{I_{NR}} D_{DI})}{2 k_B T} \right], \qquad (13)$$

Note that the unit of $R_S$ and $\alpha'$ are $\Omega$ while that of $D_{DI}$ is $\Omega \cdot A^{0.5}$. Figure 6 shows $I_R$ and $I_{NR}$ data fitted by eqs. (12) and (13). It is seen that the modified Shockley equation [eq. (12) and (13)] fits the experimental data perfectly well for both current components. The fitting parameters are $I_{S,R} = 1.3 \times 10^{-45}$ A, $I_{S,NR} = 2.3 \times 10^{-24}$ A, $R_S = 2.6\,\Omega$, $\alpha' = 4.9\,\Omega$, and $D_{DI} = 1.8\,\Omega \cdot A^{0.5}$, respectively. The inset in Fig. 6 is the schematic illustration of the equivalent circuit. The equivalent circuit basically consists of an ideal diode and a resistor. As shown in equivalent circuit, additional potential drop for each current component such as phase space filling and double injection current should be considered to account for *I-V* curve in InGaN-based LEDs.

In conclusion, to clarify the origin of high operating voltage in InGaN-based LED, we separately analyze the $I_R$ and $I_{NR}$ in combination with the Shockley diode equation, carefully. Through the analyses, we have shown that the increase in operating voltage is caused by phase space filling effect in high injection. We have also shown that the classical Shockley diode equation is insufficient to comprehensively explain the *I-V* curve of the LED devices since the transport and recombination characteristics of respective current components are basically different. Hence, we have proposed a modified Shockley equation suitable for modern LED devices. In particular, it is considered that the low degree of carrier recombination rate can lead to both the efficiency droop and the high operating voltage in InGaN-based LED devices. To remedy this, the radiative recombination rate influenced by reduced effective volume, QCSE, asymmetry carrier distribution in MQWs, and local potential fluctuation should be increased further, which will enhance the WPE in InGaN-based LEDs. We believe that characteristics of $I_R$ and $I_{NR}$ pointed out in this letter give a new insight into the electrical and optical characteristics of LED devices.

**Figures Captions**

FIG. 1. IQE characteristics of the InGaN LED sample under investigation plotted with (a) linear and (b) semi-log scale at cryogenic and room temperature, respectively.

FIG. 2. Ideality factors corresponding to $I_R$ and $I_{NR}$ as a function of applied voltage.

FIG. 3. (a) $I_R$ and (b) $I_{NR}$ as a function of applied voltage with ideal diode curves ($I_{R,ideal}$, $I_{NR,ideal}$) and additional potential drops ($\Delta V_R$, $\Delta V_{NR}$), respectively.

FIG. 4. (a) $e^{-\Delta V_{PSF}}$ as a function of $I_R$ and fitting curve of theoretical approach.

FIG. 5. $\Delta V_{NR}$ as a function of $I_{NR}$ and fitting curve of theoretical approach.

FIG. 6. $I_R$ and $I_{NR}$ vs. $V$ fitted by the modified Shockley equations for an InGaN LED sample under investigation. . The inset shows the equivalent circuit of an InGaN-based LED.

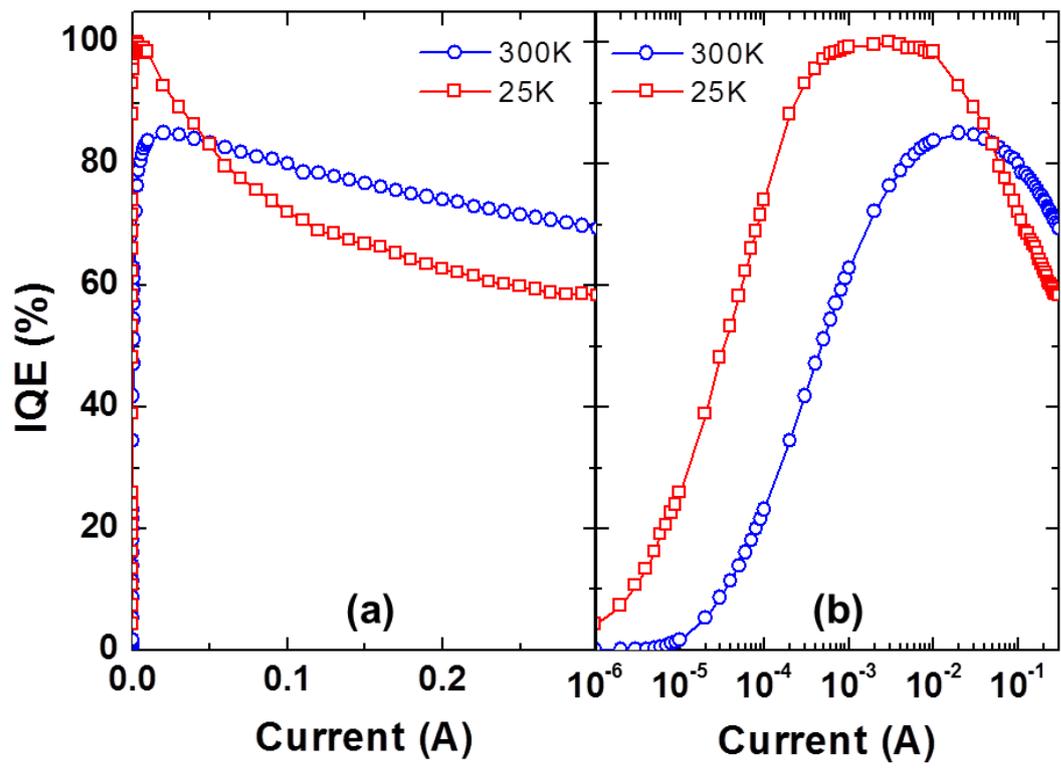

Fig. 1

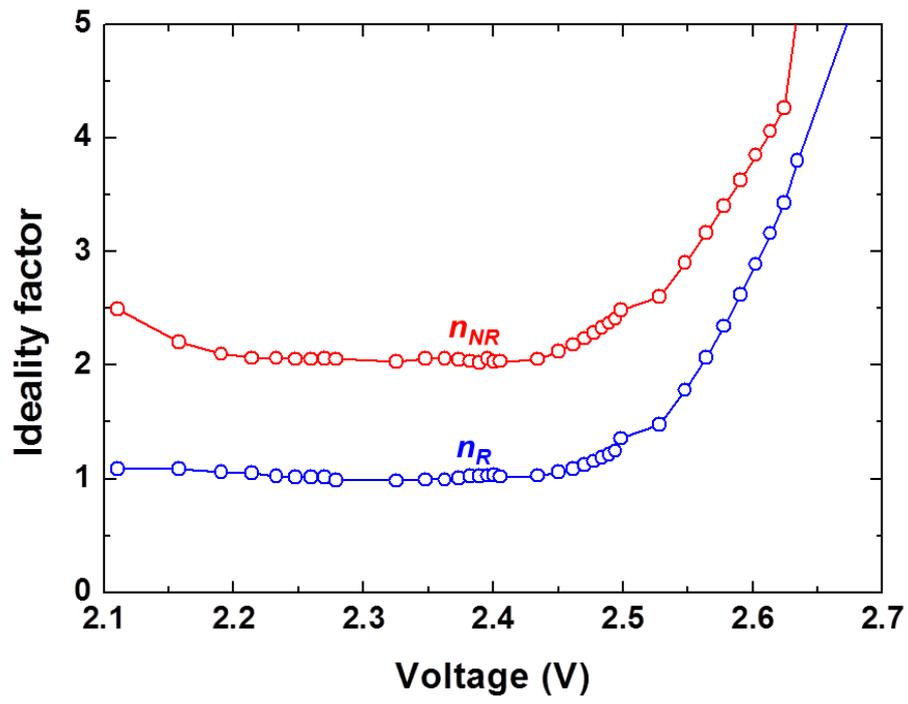

Fig. 2

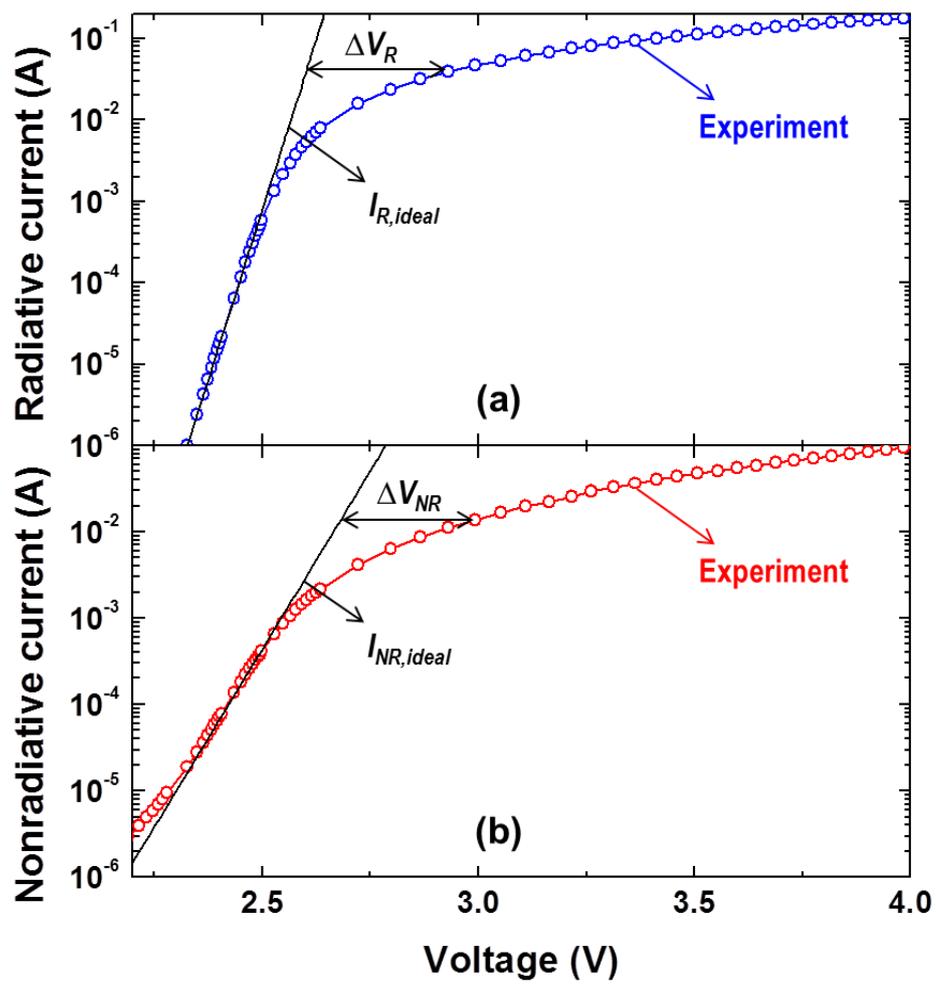

Fig. 3

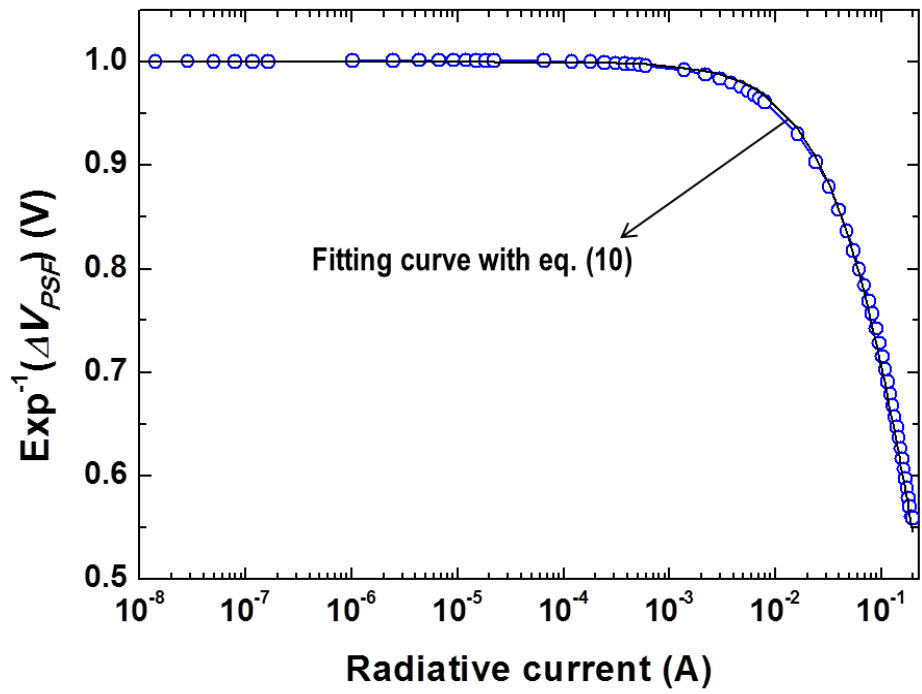

Fig. 4

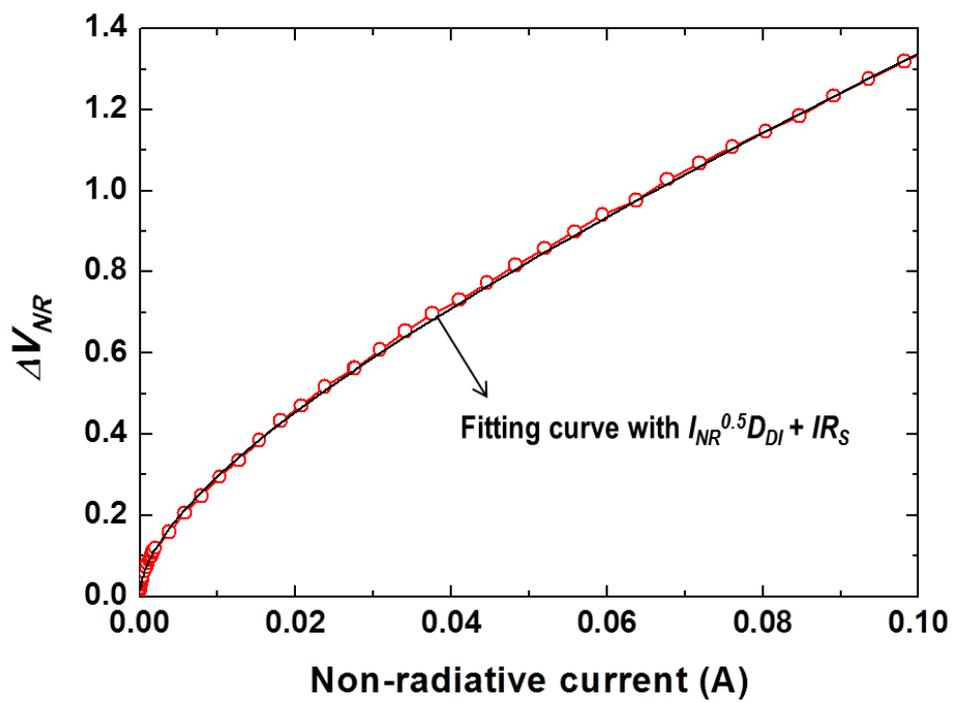

Fig. 5

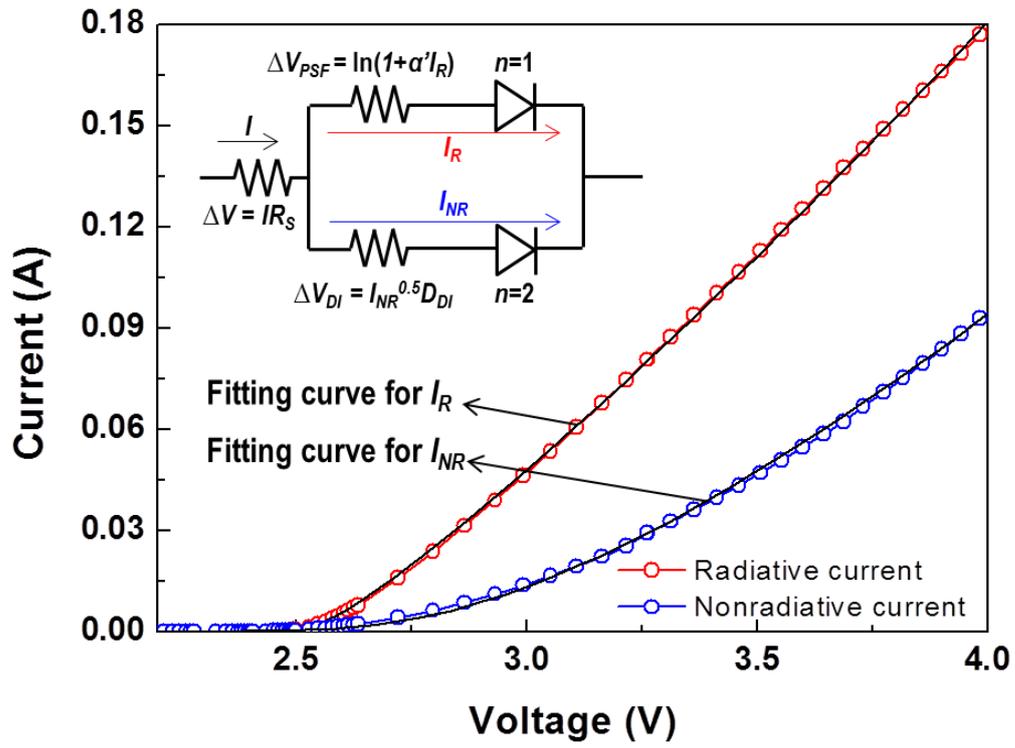

Fig. 6